\documentclass[12pt]{article}
\usepackage{amsfonts}
\usepackage{amsmath,amssymb}

\begin{document}

\title{Matter, Fields, and Reparametrization-Invariant Systems}
\author{Vesselin G. Gueorguiev\thanks{
e-mail: vesselin@phys.lsu.edu} \\
{\small \textit{Department of Physics and Astronomy,}}\\
{\small \textit{Louisiana State University, Baton Rouge, LA 70803}}}
\date{}
\maketitle

\begin{abstract}
We study reparametrization-invariant systems, mainly the relativistic
particle and its $D$-dimensional extended object generalization--$d$-brane.
The corresponding matter Lagrangians naturally contain background
interactions, like electromagnetism and gravity. For a $d$-brane that
doesn't alter the background fields, we define non-relativistic equations
assuming integral sub-manifold embedding of the $d$-brane. The mass-shell
constraint and the Klein-Gordon equation are shown to be universal when
gravity-like interaction is present. Our approach to the Dirac equation
follows Rund's technique for the algebra of the $\gamma $-matrices that
doesn't rely on the Klein-Gordon equation.
\end{abstract}

\section{Introduction}

There are two very useful methods in classical mechanics: the Hamiltonian
and the Lagrangian approach \cite{Kilmister 1967, Goldstain 1980, Gracia and
Josep, Nikitin-string theory, Carinena 1995}. The Hamiltonian formalism
gives rise to the canonical quantization, while the Lagrangian approach is
used in the path-integral quantization. Usually, in classical mechanics,
there is a transformation that relates these two approaches. However, for a
reparametrization-invariant systems there are problems when changing from
the Lagrangian to the Hamiltonian approach \cite{Goldstain 1980, Gracia and
Josep, Nikitin-string theory, Rund 1966, Lanczos 1970}. Classical mechanics
of a reparametrization-invariant system and its quantization is the topic of
the current study.

Fiber bundles provide the mathematical framework for classical mechanics,
field theory, and even quantum mechanics if viewed as a classical field
theory. When studding the structures that are important to physics, we
should also understand why one fiber bundle should be more ``physical'' than
another, why the ``physical'' base manifold seems to be a four-dimensional
Lorentzian manifold \cite{Borstnik and Nielsen, van Dam and Ng, Sachoglu
2001}, how one should construct an action integral for a given fiber bundle 
\cite{Kilmister 1967, Carinena 1995, Feynman 1965, Gerjuoy and Rau, Rivas
2001}, and how a given system may be quantized. Starting with the tangent or
cotangent bundle is natural because these bundles are related to the notion
of a classical mechanics of a point-like matter. Since our knowledge comes
from experiments that involve classical apparatus, the physically accessible
fields should be generated by matter and should couple with matter as well.
Therefore, understanding the matter Lagrangian for a classical system is
very important.

In this paper we outline some aspects of the non-relativistic, relativistic,
and a la Dirac-equation quantization of reparametrization-invariant
classical mechanics systems. In its canonical form, the matter Lagrangian
for reparametrization-invariant systems contains well known interaction
terms, such as electromagnetism and gravity. For a
reparametrization-invariant systems there are constraints among the
equations of motion which is a problem. Nevertheless, there are procedures
for quantizing such theories \cite{Nikitin-string theory, Dirac 1958a,
Teitelboim 1982, Henneaux and Teitelboim, Sundermeyer 1982}. For example,
changing coordinates $(x,v)\leftrightarrow (x,p)$ is an one problem, when $
h=pv-L\equiv 0$ is another. This $h\equiv 0$ problem is usually overcome
either by using a gauge fixing to remove the reparametrization-invariance or
by using some of the constraint equations available instead of $h$ \cite
{Nikitin-string theory}. Here, we will demonstrate another approach ( $
v\rightarrow \gamma $) which takes advantage of $h\equiv 0.$

In section two we briefly review the classical mechanics of
reparametrization-invariant $d$-branes. In the third section we argue for a
one-time-physics as an essential ingredient for a non-relativistic limit.
The fourth section is concerned with the relativistic Klein-Gordon equation,
relativistic mass-shell equation, and Dirac equation. Our conclusions and
discussions are in section five.

\section{Classical Mechanics of $d$-branes}

In this section, we briefly review our study of the geometric structures in
the classical mechanics of reparametrization-invariant systems \cite{VGG
Kiten 2002} by focusing on the relativistic charged particle and its $D$
-dimensional extended object generalization ($d$-brane). In ref. \cite{VGG
Kiten 2002} we have discussed the question: ``What is the matter Lagrangian
for a classical system?'' Starting from the assumption that there should not
be any preferred trajectory parameterization in a smooth space-time, we have
arrived at the well known and very important reparametrization-invariant
system: the charged relativistic particle. Imposing reparametrization
invariance of the action $S=\int L(x,v)\mathrm{d}\tau $ naturally leads to a
first order homogeneous functions \cite{VGG Kiten 2002}.

The Lagrangian for the charged relativistic particle corresponds to the
first two terms in a series expression of a first order homogeneous
function. When this expression is considered as a Lagrangian, we call it 
\textit{canonical form of the first order homogeneous Lagrangian}: 
\begin{equation}
L\left( \vec{x},\vec{v}\right) =\sum_{n=1}^{\infty }Q_{n}\sqrt[n]{
S_{n}\left( \vec{v},...,\vec{v}\right) }=qA_{\alpha }v^{\alpha }+m\sqrt{
g_{\alpha \beta }v^{\alpha }v^{\beta }}+...~. \label{canonical form}
\end{equation}
The choice of the canonical form is based on the assumption of one-to-one
correspondence between interaction fields $S_{n}$ and their sources \cite
{VGG Kiten 2002}. Finding a procedure, similar to the Taylor series
expansion, to extract the components of each symmetric tensor ($S_{\alpha
_{1}\alpha _{2}...\alpha _{n}}$), for a given homogeneous function of first
order, would be a significant step in our understanding of the fundamental
interactions.

Encouraged by our results, we have continued our study of
reparametrization-invariant systems by generalizing the idea to a $D$
-dimensional extended objects ($d$-branes). In doing so we have arrived at
the string theory Lagrangian (1-brane extended object) \cite{Nikitin-string
theory} and the Dirac-Nambu-Goto Lagrangian for a $d$-brane \cite{Pavsic
2001}.

The classical mechanics of a point-like particle is concerned with the
embedding $\phi :\mathbb{R}^{1}\rightarrow M$. The map $\phi $ provides
the trajectory (the word line) of the particle in the target space $M$. In
this sense, we are dealing with a $0$-brane that is a one dimensional
object. If we think of an extended object as a manifold $D$ with dimension
denoted also by $D$ ($\dim D=D=d+1$ where $d=0,1,2,...$), then we should
seek $\phi :D\rightarrow M$ such that some action integral is minimized.
From this point of view, we are dealing with embedding of a $D$-dimensional
object in to a target space $M$. If $x^{\alpha }$ denote coordinate
functions on $M$ and $z^{i}$ coordinate functions on $D$, then we can
introduce a \textit{\ generalized velocity vector} $\vec{\omega}$ with
components $\omega ^{\Gamma }$: 
\[
\omega ^{\Gamma }=\frac{\Omega ^{\Gamma }}{dz}=\frac{\partial \left(
x^{\alpha _{1}}x^{\alpha _{2}}...x^{\alpha _{D}}\right) }{\partial
(z^{1}z^{2}...z^{D})},\quad dz=dz^{1}\wedge ...\wedge dz^{D},\quad \Gamma
=1,...,\binom{\dim M}{\dim D}. 
\]
In the above expression, $\frac{\partial \left( x^{\alpha _{1}}x^{\alpha
_{2}}...x^{\alpha _{D}}\right) }{\partial (z^{1}z^{2}...z^{D})}$ represents
the Jacobian of the transformation from coordinates $\{x^{\alpha }\}$ over
the manifold $M$ to coordinates $\{z^{a}\}$ over the $d$-brane \cite{Fairlie
and Ueno}. In this notation the canonical expression for the homogeneous
Lagrangian of first order is: 
\begin{equation}
L\left( \vec{\phi},\vec{\omega}\right) =\sum_{n=1}^{\infty }\sqrt[n]{
S_{n}\left( \vec{\omega},...,\vec{\omega}\right) }=A_{\Gamma }\omega
^{\Gamma }+\sqrt{g_{\Gamma _{1}\Gamma _{2}}\omega ^{\Gamma _{1}}\omega
^{\Gamma _{2}}}+...\ . \label{canonical d-brane L}
\end{equation}
Notice that $\vec{x},\vec{v}$, and $\vec{\phi}=\vec{x}\circ \phi $ have the
same number of components when the generalized velocity $\vec{\omega}$ has $
\binom{\dim M}{\dim D}.$

From the above expressions (\ref{canonical form}) and (\ref{canonical
d-brane L}), one can see that the corresponding matter Lagrangians $(L)$, in
their canonical form, contain electromagnetic ($A$) and gravitational ($g$)
interactions, as well as interactions that are not clearly identified yet ($
S_{n},n>2$). At this stage, we have a theory with background fields since we
don't know the equations for the interaction fields $A,$ $g,$and $S_{n}$. To
complete the theory, we need to introduce actions for these interaction
fields. If one is going to study the new interaction fields $S_{n},n>2$,
then some guiding principles for writing field Lagrangians are needed.

One such principle uses the external derivative $d$, multiplication $\wedge $
, and Hodge dual $*$ operations in the external algebra $\Lambda \left(
T^{*}M\right) $ over $M$ to construct objects proportional to the volume
form over $M$. For example, for any $n$-form $(A)$ the expressions $A\wedge
*A$ and $dA\wedge *dA$ are forms proportional to the volume form.

The next important principle comes from the symmetry in the matter equation.
That is, if there is a transformation $A\rightarrow A^{\prime }$ that leaves
the matter equations unchanged, then there is no way to distinguish $A$ and $
A^{\prime }$. Thus the action for the field $A$ should obey the same gauge
symmetry. For the electromagnetic field ($A\rightarrow A^{\prime }=A+\mathrm{
d}f$) this leads to the field Lagrangian $\mathcal{L}=dA\wedge *dA=F\wedge
*F $, when for gravity it leads to the Cartan-Einstein action $S\left[
R\right] =\int R_{\alpha \beta }\wedge *(\mathrm{d}x^{\alpha }\wedge
dx^{\beta })$ \cite{VGG Kiten 2002, Adak et al 2001}. In our study we have
also found an extra term ($R^{\wedge }$) that exists only in four
dimensional theories. This term $R^{\wedge }$comes from fully
anti-symmetrized ($R_{\alpha [\beta ,\gamma ]\rho )}$) Ricci tensor $R.$

\section{Non-relativistic Limit}

Here, we briefly argue that a one-time-physics is needed to assure causality
via finite propagational speed in case of point particles. For $d $-branes
the one-time-physics reflects separation of the internal from the external
coordinates when the $d $-brane is considered as a sub-manifold of the
target space manifold $M $. The non-relativistic limit is considered to be
the case when the $d $-brane is embedded as a sub-manifold of $M$.

\subsection{Causality and Space-Time Metric Signature}

It is well known that the Einstein general relativity occurs more degree of
freedom in four and higher dimensions. We have already mentioned the $
R^{\wedge }$ term which is only possible in a four-dimensional space-time.
Another argument for 4D space-time is based on geometric and differential
structure of various brane and target spaces \cite{VGG Kiten 2002, Sachoglu
2001}. All these are reasons why the spacetime seems to be four dimensional.
Why the space-time seems to be 1+3 have been recently discussed by using
arguments a la Wigner \cite{Borstnik and Nielsen, van Dam and Ng}. However,
these arguments are deducing that the space-time is 1+3 because we only this
signature is consistent with particles with finite spin. In our opinion one
should turn this argument backwards claiming that one should observe only
particles with finite spin because the signature is 1+3.

Here we present an argument that only one-time-physics is consistent with a
finite propagational speed. Our main assumptions are: a gravity-like term $
\sqrt{g(\vec{\omega},\vec{\omega})}$ is always present in the matter
Lagrangian, and the matter Lagrangian is a real-valued; thus $g(\vec{v},\vec{
v})\geq 0.$ For simplicity, we consider the $0$-brane mechanics first.

The use of a covariant formulation allows one to select a local coordinate
system so that the metric is diagonal $(+,+,..,+,-,...-)$. If we denote the
(+) coordinates as time coordinates and the (-) as space coordinates, then
there are three essential cases:

\begin{itemize}
\item[(1)]{No time coordinates. Thus $g(\vec{v},\vec{v})=-\sum_{\alpha
}\left( v^{\alpha }\right) ^{2}<0$, which contradicts
$(g(\vec{v},\vec{v})\geq 0$).}

\item[(2)]{Two or more time coordinates. Thus $g(\vec{v},\vec{v}
)=(v^{0})^{2}+(v^{1})^{2}-\sum_{\alpha =2}^{n}\left( v^{\alpha }\right)
^{2}\Rightarrow 1+\dot{\lambda}\geq \vec{v}_{space}^{2}$.}

\item[(3)]{Only one time coordinate. Thus
$g(\vec{v},\vec{v})=(v^{0})^{2}-\sum_{
\alpha =1}^{n}\left( v^{\alpha }\right) ^{2}\Rightarrow 1\geq \vec{v}
_{space}^{2}$.}
\end{itemize}

Clearly for two or more time coordinates we don't have finite coordinate
velocity ($dx/dt $) when the coordinate time ($t $) is chosen so that $
t=x^{0}\Rightarrow $ $v^{0}=1 $ and $x^{1}=\lambda$. Only the space-time
with one time accounts for a finite velocity and thus a causal structure.

For a $d$-brane one has to assume a local coordinate frame where one
component of the generalized velocity can be set to 1 ($\omega ^{0}=1$).
This generalized velocity component is associated with the brane ``time
coordinate.'' In fact, $\omega ^{0}=1$ means that there is an integral
embedding of the $d$-brane in the target space $M$, and the image of the $d$
-brane is a sub-manifold of $M$. If the coordinates of $M$ are labeled so
that $x^{i}=z^{i},i=1,...,D$, then $x^{i}$ are internal coordinates that may
be collapsed in only one coordinate-- the ``world line'' of the $d$-brane.

\subsection{The Quantum Mechanics of a $D $-brane.}

In this section we briefly describe the gauge-fixing approach that allows
canonical quantization. This approach is mainly concerned with a choice of a
coordinate time that is used as the trajectory parameter \cite
{Nikitin-string theory, Dirac 1958, Henneaux and Teitelboim, Schwinger 1963}
. Such choice removes the reparametrization invariance of the theory.

In a local coordinate system where $\omega ^{0}=1$ and the metric is a
``one-time-metric'' we have: 
\begin{eqnarray*}
L &=&A_{\Gamma }\omega ^{\Gamma }+\sqrt{g_{\Gamma _{1}\Gamma _{2}}\omega
^{\Gamma _{1}}\omega ^{\Gamma _{2}}}+...+\sqrt[m]{S_{m}\left( \vec{\omega}
,...,\vec{\omega}\right) }\rightarrow \\
&\rightarrow &A_{0}+A_{i}\omega ^{i}+\sqrt{1-g_{ii}\omega ^{i}\omega ^{i}}
+...\approx A_{0}+A_{i}\omega ^{i}+1-\frac{1}{2}g_{ii}\omega ^{i}\omega
^{i}+....
\end{eqnarray*}
Thus the Hamiltonian function is not zero anymore, so we can do canonical
quantization, and the Hilbert space consists of the functions $\Psi \left(
x\right) \rightarrow \Psi \left( z,\tilde{x}\right) $ where $\tilde{x}
=x^{i},i=D+1,...,m$. The brane coordinates $z$ shall be treated as $t$ in
quantum mechanics in the sense that the scalar product should be an integral
over the space coordinates $\tilde{x}$.

\section{Relativistic Equations for Matter}

Even though canonical quantization can be applied after a gauge fixing, one
is not usually happy with this situation because the covariance of the
theory is lost and time is a privileged coordinate. In general, there are
well developed procedures for covariant quantization \cite{Nikitin-string
theory, Dirac 1958a, Teitelboim 1982, Henneaux and Teitelboim, Sundermeyer
1982}. However, we are not going to discuss these methods. Instead, we will
employ a different quantization strategy. In this section we discuss the
mass-shell constraint, the Klein-Gordon equation and the Dirac equation \cite
{Schweber 1961} for $d$-branes.

\subsection{The Mass-shell and Klein-Gordon Equation}

Since the functional form of the canonical Lagrangian is the same for any $d$
-brane, we use $v$, but it could be $\omega $ as well. We define the
momentum $p$ and generalized momentum $\pi $ for our canonical Lagrangian as
follow: 
\begin{eqnarray*}
p_{\Gamma } &=&\frac{\delta L\left( \phi ,\omega \right) }{\delta \omega
^{\Gamma }}=eA_{\Gamma }+m\frac{g_{\Gamma \Sigma }\omega ^{\Sigma }}{\sqrt{
g\left( \vec{\omega},\vec{\omega}\right) }}+...+\frac{S_{\Gamma \Sigma
_{1}...\Sigma _{n}}\omega ^{\Sigma _{1}}...\omega ^{\Sigma _{n}}}{\left(
S\left( \omega ,...,\omega \right) \right) ^{1-1/n}}+..., \\
\pi _{\alpha } &=&p_{\alpha }-eA_{\alpha }-...\frac{S_{\alpha \beta
_{1}...\beta _{n}}v^{\beta _{1}}...v^{\beta _{n}}}{\left( S\left(
v,...,v\right) \right) ^{n/\left( n+1\right) }}...=m\frac{g_{\alpha \beta
}v^{\beta }}{\sqrt{g\left( \vec{v},\vec{v}\right) }}.
\end{eqnarray*}
In the second equation we have used $v$ instead of $\omega $ for simplicity.
Notice that this generalized momentum ($\pi $) is consistent with the usual
quantum mechanical procedure $p\rightarrow p-eA$ that is used in Yang-Mills
theories, as well as with the usual GR expression $p_{\alpha }=mg_{\alpha
\beta }v^{\beta }$. Now it is easy to recognize the mass-shell constraint as
a mathematical identity: 
\[
\frac{\vec{v}}{\sqrt{\vec{v}^{2}}}\cdot \frac{\vec{v}}{\sqrt{\vec{v}^{2}}}
=1\Rightarrow \pi _{\alpha }\pi ^{\alpha }=m^{2}\quad \Rightarrow \quad
\left( \vec{p}-e\vec{A}-\vec{S}_{3}\left( v\right) -\vec{S}_{4}\left(
v\right) -...\right) ^{2}\Psi =m^{2}\Psi . 
\]
Notice that ``gravity'' as represented by the metric is gone, while the
Klein-Gordon equation appears. The $v$ dependence in the $S$ terms reminds
us about the problem related to the change of coordinates $(x,v)\rightarrow
(x,p)$. So, at this stage we may proceed with the Klein-Gordon equation, if
we wish.

\subsection{Rund's Approach to the $\gamma $-matrices}

An interesting approach to the Dirac equation has been suggested by H. Rund 
\cite{Rund 1966}. The idea uses the Heisenberg picture ($-i\hbar \frac{dZ}{
d\tau }=[H,Z]$), Hamiltonian linear in the momentum ($H=\gamma ^{\alpha
}p_{\alpha }$), and principle group $G$ with generators $X_{i}$ that close a
Lie algebra ($[X_{i},X_{j}]=C_{ij}^{k}X_{k}$). To have the Hamiltonian $H$
invariant under $G$-transformations, the $\gamma $ objects should transform
appropriately $[X_{i},\gamma ^{\alpha }]=\left( \rho \left( X_{i}\right)
\right) _{\beta }^{\alpha }\gamma ^{\beta }$. The next ansatz is the
important one: $X_{i}=\left( x_{i}\right) _{\alpha \beta }\gamma ^{\alpha
}\gamma ^{\beta }.$

By using this ansatz in $[X_{i},X_{j}]=C_{ij}^{k}X_{k}$, one writes $
[X_{i},(x_{j})_{\alpha \beta }\gamma ^{\alpha }\gamma ^{\beta
}]=C_{ij}^{k}(x_{k})_{\alpha \beta }\gamma ^{\alpha }\gamma ^{\beta }$ and
solves for $(x_{j})_{\alpha \beta }$. In order for the linear system of
equations to have a solution, additional algebraic properties on the $\gamma 
$ matrices are required. For the Lorentz group, this procedure gives $
\Lambda ^{\alpha \beta }=\frac{1}{2}\gamma ^{\alpha }\gamma ^{\beta }$ with $
\{\gamma ^{\alpha },\gamma ^{\beta }\}=2\delta ^{\alpha \beta }$. Notice
that in $H=\gamma ^{\alpha }p_{\alpha }$ the momentum transforms according
to the fundamental representation, and thus the $\gamma $-vector should
transform as the conjugate one, so that $H$ stays a scalar. The ansatz means
that we are constructing the adjoint representation, which is the Lie
algebra itself, by coupling two fundamental representations.

\subsection{Dirac Equation from H=0}

Since we want $\gamma $ and $p $ to transform as vectors, it is clear that $
p $ should be a covariant derivative, but what is its structure? Consider a
homogeneous Lagrangian that can be written as $L\left(\phi,\omega \right)
=\omega ^{\Gamma}p_{\Gamma}=\omega ^{\Gamma}\partial L\left(\phi,\omega
\right) /\partial \omega ^{\Gamma} $ with a Hamiltonian function that is
identically zero: $h=\omega ^{\Gamma}\partial L\left(\phi,\omega \right)
/\partial \omega ^{\Gamma}-L\left(\phi,\omega \right) \equiv 0$. Notice that 
$\omega ^{\Gamma} $ is the determinant of a matrix (the Jacobian of a
transformation \cite{Fairlie and Ueno}); thus $\omega ^{\Gamma}\rightarrow
\gamma ^{\Gamma} $ seems an interesting option for quantization. Even more,
for the Dirac theory we know that $\gamma ^{\alpha} $ are the `velocities' ($
dx/d\tau =\partial H/\partial p $).

If we quantize using ($h\rightarrow H$), then for the space of functions we
should have: $H\Psi =0$. By applying $\omega ^{\Gamma }\rightarrow \gamma
^{\Gamma }$, which means that the (generalized) velocity is considered as a
vector with non-commutative components, we have $\left( \gamma ^{\Gamma
}p_{\Gamma }-L\left( \phi ,\gamma \right) \right) \Psi =0$. For a 0-brane,
using the canonical form of the Lagrangian (\ref{canonical form}) and the
algebra of the $\gamma $ matrices following Run's approach, as described in
the previous section, we have: 
\begin{eqnarray*}
H &=&\gamma ^{\alpha }p_{\alpha }-L\left( \phi ,\gamma \right) =\gamma
^{\alpha }p_{\alpha }-eA_{\alpha }\gamma ^{\alpha }-m\sqrt{g_{\alpha \beta
}\gamma ^{\alpha }\gamma ^{\beta }}-...\sqrt[m]{S_{m}\left( \vec{\gamma},...,
\vec{\gamma}\right) }, \\
&\rightarrow &\gamma ^{\alpha }p_{\alpha }-eA_{\alpha }\gamma ^{\alpha
}-m-...\sqrt[2m]{S_{2m}g^{m}}-...\sqrt[2n+1]{S_{2n+1}g^{n/2}\gamma }....
\end{eqnarray*}
Since $g_{\alpha \beta }$ is a symmetric tensor, then $g_{\alpha \beta
}\gamma ^{\alpha }\gamma ^{\beta }\sim g_{\alpha \beta }\{\gamma ^{\alpha
},\gamma ^{\beta }\}\sim g_{\alpha \beta }g^{\alpha \beta }\sim 1$.
Therefore, gravity seems to leave the picture again. The symmetric structure
of the extra terms $S_{m}$ can be used to reintroduce $g$ by using $\{\gamma
^{\alpha },\gamma ^{\beta }\}\sim g^{\alpha \beta }$ and to reduce the
powers of $\gamma $. Thus the high even terms contribute to the mass $m$,
making it variable with $\vec{x}$ \cite{Bekenstein 1993}.

\section{Conclusions and Discussions}

In summary, we have discussed the structure of the matter Lagrangian ($L$)
for extended objects. Imposing reparametrization invariance of the action $S$
naturally leads to a first order homogeneous Lagrangian. In its canonical
form, $L$ contains electromagnetic and gravitational interactions, as well
as interactions that are not clearly identified yet.

The non-relativistic limit for a $d$-brane has been defined as those
coordinates where the brane is an integral sub-manifold of the target space.
This gauge can be used to remove reparametrization invariance of the action $
S$ and make the Hamiltonian function suitable for canonical quantization.
For the 0-brane (the relativistic particle), this also has a clear physical
interpretation associated with localization and finite propagational speed.

The existence of a mass-shell constraint is universal. It is essentially due
to the gravitational (quadratic in velocities) type interaction in the
Lagrangian and leads to a Klein-Gordon equation. Although the Klein-Gordon
equation can be defined, it is not the only way to introduce the algebra of
the $\gamma $-matrices needed for the Dirac equation. The algebraic
properties of the $\gamma $-matrices may be derived using the Lie group
structure of the coordinate bundle; these properties are closely related to
the corresponding metric tensor $g^{\alpha \beta }=\left\{ \gamma ^{\alpha
},\gamma ^{\beta }\right\} $ and may restrict the number of terms in the
Lagrangian. Once the algebraic properties of the $\gamma $-matrices are
defined, one can use $v\rightarrow \gamma $ quantization in the Hamiltonian
function $h=pv-L\left( x,v\right) $ to obtain the Dirac equation.

\textbf{Acknowledgments.} The author acknowledges helpful discussions with
Professors R. Hymaker, L. Smolinsky, A. R. P. Rau, R. F. O'Connell, P. Kirk,
J. Pullin, C. Torre, J. Baez, P. Al. Nikolov, E. M. Prodanov, G. Dunne, and
L. I. Gould. The author is also thankful to Professor J. P. Draayer at the
LSU Department of Physics and Astronomy, Dr. Joe Abraham at the LSU Writing
Center, and the organizers of the 4th Conference on Geometry, Integrability
and Quantization in Sts. Constantine and Elena resort, Bulgaria, who
provided the opportunity for this research and its publication. The author
acknowledges the financial support from the Department of Physics and
Astronomy and the Graduate School at the Louisiana State University, the U.
S. National Science Foundation support under Grant No. PHY-9970769 and
Cooperative Agreement No. EPS-9720652 that includes matching from the
Louisiana Board of Regents Support Fund.

\end{document}